\documentclass[aps, twocolumn, superscriptaddress, floatfix, nofootinbib, amsmath, amssymb, preprintnumbers]{revtex4-1}

\usepackage{dcolumn}
\usepackage{verbatim}

\usepackage{breakurl}

\usepackage{lmodern}

\pdfoutput=1
\usepackage{graphicx}
\usepackage{bm}
\usepackage{amsmath}
\usepackage{amsfonts}
\usepackage{amssymb}
\usepackage{multirow}
\usepackage[colorlinks,linktocpage,linkcolor=cyan,citecolor=cyan]{hyperref}
\usepackage{adjustbox}
\usepackage{array}
\usepackage[capitalise]{cleveref}
\usepackage{acro}
\usepackage[utf8]{inputenc}
\usepackage{CJKutf8}
\usepackage{amssymb}
\usepackage{subfigure}

\usepackage{tikz}
\usepackage{pgfplots}
\pgfplotsset{compat=newest,every axis plot/.append style={line width=1pt}}

\crefname{figure}{Fig.}{Figs.}
\Crefname{figure}{Fig.}{Figs.}
\def\({\left(}
\def\){\right)}
\def\[{\left[}
\def\]{\right]}

\newcommand{\be}{{\begin{eqnarray}}}
\newcommand{\ee}{{\end{eqnarray}}}

\newcommand{\overbar}[1]{\mkern 1.5mu\overline{\mkern-1.5mu#1\mkern-1.5mu}\mkern 1.5mu}

\newcommand{\fnl}{f_\mathrm{NL}}
\newcommand{\Fnl}{F_\mathrm{NL}}

\newcommand{\bk}{\mathbf{k}}
\newcommand{\bq}{\mathbf{q}}

\newcommand{\bx}{\mathbf{x}}
\newcommand{\bn}{\mathbf{n}}

\newcommand{\ud}{\mathrm{d}}

\newcommand{\uGW}{\mathrm{gw}}
\newcommand{\uc}{\mathrm{crit}}
\newcommand{\Beq}{\begin{align}}
\newcommand{\Eeq}{\end{align}}

\DeclareAcronym{SW}{
  short = SW ,
  long = Sachs-Wolfe ,
  short-plural =  ,
}
\DeclareAcronym{BH}{
  short = BH ,
  long = black hole ,
  short-plural = s ,
}
\DeclareAcronym{SNR}{
  short = SNR ,
  long = signal-to-noise ratio ,
  short-plural = s ,
}
\DeclareAcronym{IMRPPv2}{
  short = ,
  long = {\normalsize IMRP}{\footnotesize HENOM}{\normalsize P}v2 ,
  short-plural = ,
}
\DeclareAcronym{SFR}{
  short = SFR ,
  long = star formation rate ,
  short-plural =  ,
}
\DeclareAcronym{IMR}{
  short = IMR ,
  long = inspiral-merger-ringdown ,
  short-plural =  ,
}
\DeclareAcronym{ABH}{
	short = ABH ,
	long  = astrophysical black hole,
  short-plural = s ,
}
\DeclareAcronym{GW}{
  short = GW ,
  long = gravitational wave ,
  short-plural = s ,
}
\DeclareAcronym{SIGW}{
  short = SIGW ,
  long = scalar-induced gravitational wave ,
  short-plural = s ,
}
\DeclareAcronym{GWB}{
  short = GWB ,
  long = gravitational-wave background ,
  short-plural = s ,
}
\DeclareAcronym{CBC}{
  short = CBC ,
  long = compact binary coalescence ,
  short-plural = s ,
}
\DeclareAcronym{BBH}{
  short = BBH ,
  long = binary black hole ,
  short-plural = s ,
}
\DeclareAcronym{PBH}{
  short = PBH ,
  long = primordial black hole ,
  short-plural = s ,
}
\DeclareAcronym{LIGO}{
  short =LIGO ,
  long = Laser Interferometer Gravitational-Wave Observatory ,
  short-plural = ,
}
\DeclareAcronym{LVK}{
  short = LVK ,
  long = {Advanced LIGO, Virgo and KAGRA} ,
  short-plural = ,
}
\DeclareAcronym{ET}{
	short = ET ,
	long  = Einstein Telescope,
  short-plural =  ,
}
\DeclareAcronym{CE}{
	short = CE ,
	long  = Cosmic Explorer,
  short-plural =  ,
}
\DeclareAcronym{LISA}{
	short = LISA ,
	long  = Laser Interferometer Space Antenna,
  short-plural =  ,
}
\DeclareAcronym{BBO}{
	short = BBO ,
	long  = big bang observer,
  short-plural =  ,
}
\DeclareAcronym{DECIGO}{
	short = DECIGO ,
	long  = Deci-hertz Interferometer Gravitational wave Observatory,
  short-plural =  ,
}
\DeclareAcronym{SKA}{
	short = SKA ,
	long  = Square Kilometre Array,
  short-plural =  ,
}
\DeclareAcronym{PTA}{
  short = PTA ,
  long = pulsar timing array ,
  short-plural = s ,
}
\DeclareAcronym{FRW}{
  short = FRW ,
  long = Friedman-Robertson-Walker ,
  short-plural =  ,
}
\DeclareAcronym{CMB}{
  short = CMB ,
  long = cosmic microwave background ,
  short-plural =  ,
}
\DeclareAcronym{RD}{
  short = RD,
  long  = radiation-dominated ,
  short-plural =  ,
}
\DeclareAcronym{MD}{
  short = MD,
  long  = matter-dominated ,
  short-plural =  ,
}
\DeclareAcronym{HD}{
  short = HD,
  long  = Hellings-Downs ,
  short-plural =  ,
}
\DeclareAcronym{SMBH}{
  short = SMBH ,
  long  = supper-massive black hole ,
  short-plural = s ,
}
\DeclareAcronym{SGWB}{
  short = SGWB ,
  long  = stochastic gravitational-wave background ,
  short-plural = s ,
}
\DeclareAcronym{NG15}{
  short = NG15 ,
  long  = NANOGrav 15-year ,
  short-plural =  ,
}
\DeclareAcronym{PSD}{
  short = PSD ,
  long  = power spectral density ,
  short-plural = s ,
}
\DeclareAcronym{PDF}{
  short = PDF ,
  long  = probability distribution function ,
  short-plural = s ,
}

\begin{document}

\preprint{APS/123-QED}

\title{Implications of Pulsar Timing Array Data for Scalar-Induced Gravitational Waves and Primordial Black Holes: Primordial Non-Gaussianity $f_{\mathrm{NL}}$ Considered}

\author{Sai Wang}
\affiliation{Theoretical Physics Division, Institute of High Energy Physics, Chinese Academy of Sciences, Beijing 100049, People's Republic of China}

\author{Zhi-Chao Zhao}
\email{Corresponding author: zhaozc@cau.edu.cn}
\affiliation{Department of Applied Physics, College of Science, China Agricultural University,
Qinghua East Road, Beijing 100083, People's Republic of China}

\author{Jun-Peng Li}
\affiliation{Theoretical Physics Division, Institute of High Energy Physics, Chinese Academy of Sciences, Beijing 100049, People's Republic of China}
\affiliation{School of Physical Sciences, University of Chinese Academy of Sciences, Beijing 100049, People's Republic of China}

\author{Qing-Hua Zhu}
\affiliation{CAS Key Laboratory of Theoretical Physics, Institute of Theoretical Physics,
Chinese Academy of Sciences, Beijing 100190, People's Republic of China}


\begin{abstract}
Multiple pulsar-timing-array collaborations have reported strong evidence for the existence of a gravitational-wave background. We study physical implications of this signal for cosmology, assuming that it is attributed to scalar-induced gravitational waves. By incorporating primordial non-Gaussianity $f_{\mathrm{NL}}$, we specifically examine the nature of primordial curvature perturbations and primordial black holes. We find that the signal allows for a primordial non-Gaussianity $f_{\mathrm{NL}}$ in the range of $-4.1\lesssim f_{\mathrm{NL}} \lesssim 4.1$ (68\% confidence intervals) and a mass range for primordial black holes $m_{\mathrm{pbh}}$ spanning from $\sim10^{-5}M_{\odot}$ to $\sim10^{-2}M_{\odot}$. Furthermore, we find that the signal favors a negative non-Gaussianity, which can suppress the abundance of primordial black holes. We also demonstrate that the anisotropies of scalar-induced gravitational waves serve as a powerful tool to probe the non-Gaussianity $f_{\mathrm{NL}}$. We conduct a comprehensive analysis of the angular power spectrum within the nano-Hertz band. Looking ahead, we anticipate that future projects, such as the Square Kilometre Array, will have the potential to measure these anisotropies and provide further insights into the primordial universe.

\end{abstract}

\maketitle


\section{Introduction}

Multiple collaborations in the field of \ac{PTA} observations have presented strong evidence for a signal exhibiting correlations consistent with a stochastic \ac{GWB} \cite{Xu:2023wog,Antoniadis:2023ott,NANOGrav:2023gor,Reardon:2023gzh}. 
The strain has been measured to be on the order of $10^{-15}$ at a pivot frequency of $1~\mathrm{yr}^{-1}$. 
Though this \ac{GWB} aligns with expectations from astrophysical sources, specifically inspiraling \ac{SMBH} binaries \cite{NANOGrav:2023hfp}, it is important to note that the current datasets do not rule out the possibility of cosmological origins or other exotic astrophysical sources, which have been explored in collaborative accompanying papers \cite{Antoniadis:2023xlr,NANOGrav:2023hvm}. 
Notably, several cosmological models have demonstrated superior fits to the signal compared to the \ac{SMBH}-binary interpretation. 
If these alternative models are confirmed in the future, they may provide compelling evidence for new physics.

In this study, our focus lies on the cosmological interpretation of the signal, specifically the existence of \acp{SIGW} \cite{Ananda:2006af,Baumann:2007zm,Mollerach:2003nq,Assadullahi:2009jc,Espinosa:2018eve,Kohri:2018awv}. 
This possibility had been used for interpreting the NANOGrav 12.5year dataset \cite{NANOGrav:2020bcs} in Refs.~\cite{DeLuca:2020agl,Vaskonen:2020lbd,Kohri:2020qqd,Domenech:2020ers,Atal:2020yic,Yi:2022ymw,Zhao:2022kvz,Dandoy:2023jot,Cai:2021wzd,Inomata:2020xad}. 
It was recently revisited by the \ac{PTA} collaborations \cite{Antoniadis:2023xlr,NANOGrav:2023hvm}, but the statistics of primordial curvature perturbations was assumed to be Gaussian. 
However, it was demonstrated that primordial non-Gaussianity $\fnl$ significantly contributes to the energy density of \acp{SIGW} \cite{Garcia-Bellido:2017aan,Domenech:2017ems,Cai:2018dig,Unal:2018yaa,Yuan:2020iwf,Atal:2021jyo,Adshead:2021hnm,Ragavendra:2021qdu,Li:2023qua}. 
This indicates noteworthy modifications to the energy-density spectrum, which is crucial for the data analysis of \ac{PTA} datasets. 
On the other hand, it has been shown that primordial non-Gaussianity $\fnl$ could generate initial inhomogeneities in \acp{SIGW}, leading to anisotropies characterized by the angular power spectrum \cite{Li:2023qua}. 
Related studies can be found in Refs.~\cite{Bartolo:2019zvb,ValbusaDallArmi:2020ifo,Dimastrogiovanni:2021mfs,Schulze:2023ich,LISACosmologyWorkingGroup:2022kbp,LISACosmologyWorkingGroup:2022jok,Unal:2020mts,Malhotra:2020ket,Carr:2020gox}. 
Our analysis will encompass a comprehensive examination of the angular power spectrum within the \ac{PTA} band. 
Moreover, this spectrum is capable of breaking the degeneracies among model parameters, particularly leading to possible determination of $\fnl$, and playing a crucial role in distinguishing between different sources of \ac{GWB}. 
Therefore, by interpreting the signal as originating from \acp{SIGW}, we aim to study physical implications of \ac{PTA} datasets for the nature of primordial curvature perturbations, including their power spectrum and angular power spectrum.

We will also study implications of the aforementioned results for scenarios involving formation of \acp{PBH}, which was accompanied by the production of \acp{SIGW}. 
Enhanced primordial curvature perturbations not only lead to formation of \acp{PBH} through gravitational collapse \cite{Hawking:1971ei}, but also produce \ac{GWB} via nonlinear mode-couplings. 
The study of \acp{SIGW} thus allows us to explore the \ac{PBH} scenarios \cite{Bugaev:2009zh,Saito:2009jt,Wang:2019kaf,Kapadia:2020pnr,Chen:2019xse,Papanikolaou:2022chm,Papanikolaou:2020qtd,Madge:2023cak}. 
Related works analyzing observational datasets can be found in Refs.~\cite{Zhao:2022kvz,Dandoy:2023jot,Yi:2022ymw,NANOGrav:2023hvm,Antoniadis:2023xlr,Romero-Rodriguez:2021aws,Kapadia:2020pnr}, and influence of primordial non-Gaussianity on the mass function of \acp{PBH} was also studied \cite{Bullock:1996at,Byrnes:2012yx,Young:2013oia,Franciolini:2018vbk,Passaglia:2018ixg,Atal:2018neu,Atal:2019cdz,Taoso:2021uvl,Meng:2022ixx,Chen:2023lou,Kawaguchi:2023mgk,Fu:2020lob,Inomata:2020xad,Young:2014ana,Choudhury:2023kdb}. 
Taking $\fnl$ into account in this work, we will reinterpret the constraints on power spectrum as constraints on the mass range of \acp{PBH}.

The remaining context of this paper is arranged as follows. 
In \cref{sec:edfs}, we will provide a brief summary of the homogeneous and isotropic component of \acp{SIGW}. 
In \cref{sec:pps}, we will show implications of the \ac{PTA} data for the power spectrum of primordial curvature perturbations and then for the mass function of \acp{PBH}. 
In \cref{sec:asigw}, we will study the inhomogeneous and anisotropic component of \acp{SIGW} and show the corresponding angular power spectrum in \ac{PTA} band. 
In \cref{sec:cd}, we make concluding remarks and discussions.

\section{SIGW energy-density fraction spectrum}\label{sec:edfs}

In this section, we show a brief but self-consistent summary of the main results of the energy-density fraction spectrum in a framework of \ac{SIGW} theory.

For the homogeneous and isotropic component of a \ac{GWB}, the energy-density fraction spectrum is defined as $\bar{\Omega}_{\uGW} (\eta,q) = \bar{\rho}_{\uGW} (\eta,q) / \rho_\uc(\eta)$ \cite{Maggiore:1999vm}, where $q$ represents the wavenumber, $\rho_\uc$ denotes the critical energy density of the universe at conformal time $\eta$, and the overbar signifies quantities at the background level. 
This definition implies that $\int \bar{\rho}_{\uGW} (\eta,q) \ud \ln q$ corresponds to the energy-density fraction of \ac{GWB} \cite{Maggiore:1999vm}. 
The spectrum can be formally expressed as $\bar{\rho}_{\uGW}(\eta,q) \sim \langle h_{ij,l} h_{ij,l} \rangle$, where $h_{ij}(\eta,\bq)$ represents the strain with wavevector $\bq$, and the angle brackets denote an ensemble average.
For subhorizon-scale \acp{SIGW}, we have $h_{ij}\sim\zeta^2$, leading to $\bar{\Omega}_\uGW(\eta,\bq)\sim\langle\zeta^4\rangle$ \cite{Ananda:2006af,Baumann:2007zm}, where $\zeta(\bq)$ represents curvature perturbations in the early universe.
In the case of primordial Gaussianity, semi-analytic formulas for $\bar{\Omega}_\uGW(\eta,\bq)$ were derived in Refs.~\cite{Espinosa:2018eve,Kohri:2018awv}, with earlier relevant works in Refs.~\cite{Ananda:2006af,Baumann:2007zm}. 
However, in the presence of primordial non-Gaussianity $\fnl$, there are not such semi-analytic formulas. 
Recent literature provided relevant studies on this topic \cite{Garcia-Bellido:2017aan,Domenech:2017ems,Cai:2018dig,Unal:2018yaa,Yuan:2020iwf,Atal:2021jyo,Ragavendra:2021qdu,Adshead:2021hnm,Garcia-Saenz:2022tzu,Li:2023qua}.
In this work, we adopt the conventions established in our previous study \cite{Li:2023qua}.

To quantify contributions of $\fnl$ to the energy density, we express the primordial curvature perturbations $\zeta$ in terms of their Gaussian components $\zeta_g$, i.e., \cite{Komatsu:2001rj} 
\begin{equation}
\label{eq:fnl-def-k} 
\zeta (\bq) = \zeta_g(\bq) + \frac{3}{5}\fnl \int \frac{\ud^3 \bk}{(2\pi)^{3/2}} \zeta_g(\bk) \zeta_g(\bq-\bk)\ . \end{equation} 
Here, $\fnl$ represents the non-linear parameter that characterizes the local-type primordial non-Gaussianity. 
To simplify the subsequent analytic formulae, we introduce a new quantity as follows  
\begin{equation}
    \Fnl=\frac{3}{5}\fnl \ .
\end{equation}
It is worth noting that perturbation theory requires the condition $A_S\Fnl^2<1$, where $A_S$ will be defined later.
We define the dimensionless power spectrum of $\zeta_g$ as  
\begin{equation} \langle \zeta_g (\bq) \zeta_g (\bq') \rangle = \delta^{(3)} (\bq+\bq') \frac{2 \pi^2}{q^3} \Delta^2_g (q) \ . \end{equation} 
In this work, we assume that $\Delta^2_g(q)$ follows a log-normal distribution with respect to $\ln q$ \cite{Pi:2020otn,Adshead:2021hnm,Zhao:2022kvz,Dimastrogiovanni:2022eir,Cang:2022jyc} 
\begin{equation}\label{eq:Lognormal} \Delta^2_g (q) = \frac{A_S}{\sqrt{2\pi\sigma^2}}\exp\left[-\frac{\ln^2 (q/q_\ast)}{2 \sigma^2}\right]\ , \end{equation} 
where $A_S$ represents the spectral amplitude at the spectral peak wavenumber $q_\ast$, and $\sigma$ denotes the standard deviation that characterizes the width of the spectrum.
The wavenumber $q$ is straightforwardly converted into the frequency $\nu$, namely, $q=2\pi\nu$.

\begin{figure}
    \centering
    \includegraphics[width =1 \columnwidth]{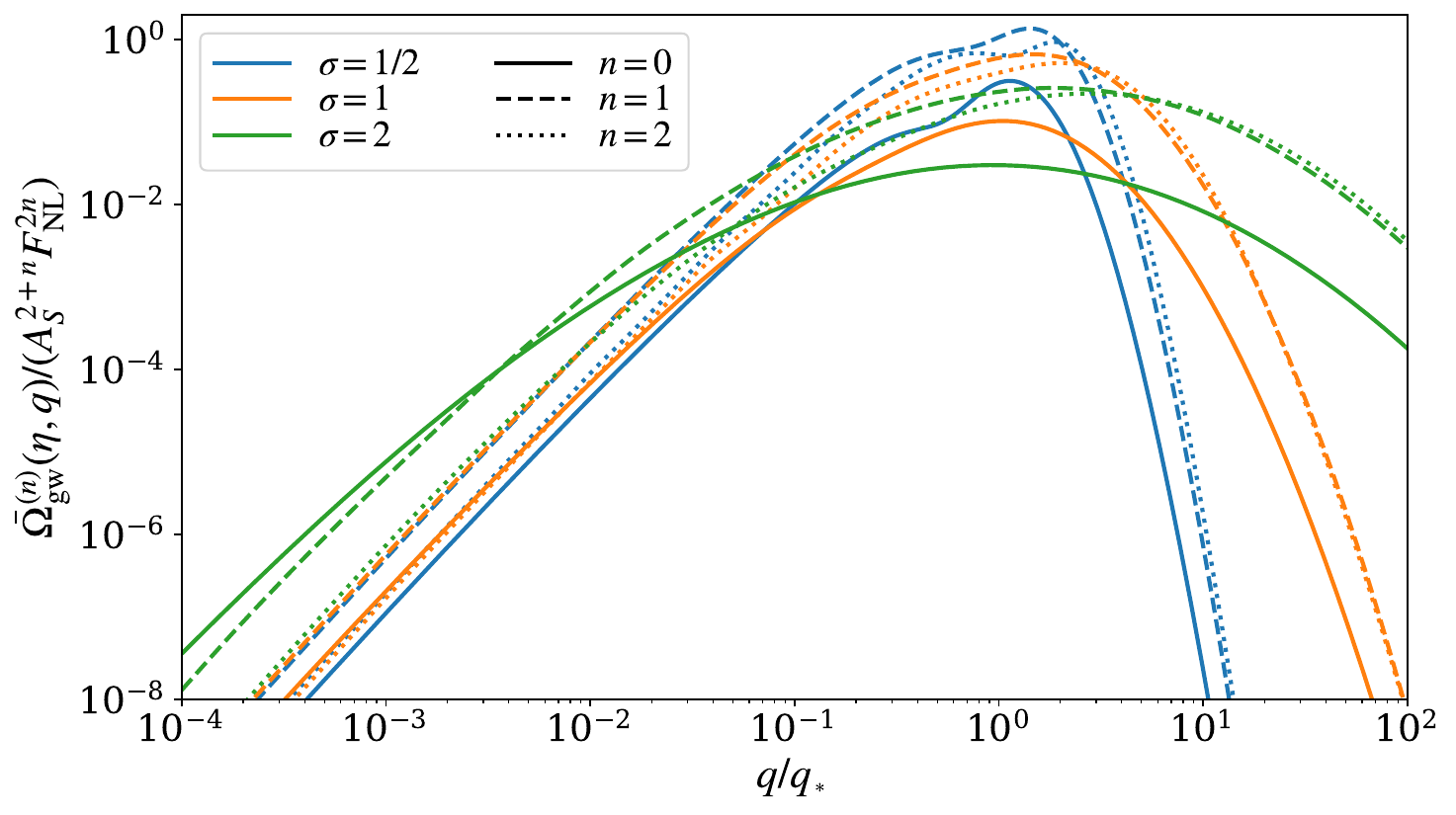}
    \caption{Unscaled (or equivalently, $A_S=1$ and $\Fnl=1$) contributions to the energy-density fraction spectrum of \acp{SIGW}. We display $\sigma=1/2,1,2$ and produce this figure by using the original data of Ref.~\cite{Li:2023qua}. }
    \label{fig:unscaled_omega}
\end{figure}

Through a detailed derivation process based on Wick's theorem, we can decompose $\bar{\Omega}_\uGW\sim\langle\zeta^4\rangle$ into three components depending on the power of $\fnl$. 
However, the complete derivations have been simplified by employing a Feynman-like diagrammatic approach \cite{Garcia-Bellido:2017aan,Unal:2018yaa,Atal:2021jyo,Ragavendra:2021qdu,Adshead:2021hnm,Li:2023qua}. 
Here, we present the final results 
\begin{equation}\label{eq:Omegabar-total} \bar{\Omega}_{\uGW} (\eta,q) = \bar{\Omega}_{\uGW}^{(0)} (\eta,q) + \bar{\Omega}_{\uGW}^{(1)} (\eta,q) + \bar{\Omega}_{\uGW}^{(2)} (\eta,q)\ . 
\end{equation}
where we provide the analytic expressions for $\bar{\Omega}_\uGW^{(n)}$, which are proportional to $A_S^2 (A_S \Fnl^2)^n$ with $n=0,1,2$, in \cref{sec:formula}. 
They were evaluated using the \texttt{vegas} package \cite{Lepage:2020tgj}, while their numerical results for $\sigma=1/2,1,2$ are reproduced in Fig.~\ref{fig:unscaled_omega}. 
Specifically, $\bar{\Omega}_\uGW^{(0)}$ corresponds to the energy-density fraction spectrum in the case of Gaussianity, while $\bar{\Omega}_\uGW^{(1)}$ and $\bar{\Omega}_\uGW^{(2)}$ fully describe the contributions of local-type primordial non-Gaussianity $\fnl$.

\begin{figure}
    \centering
    \includegraphics[width =1\columnwidth]{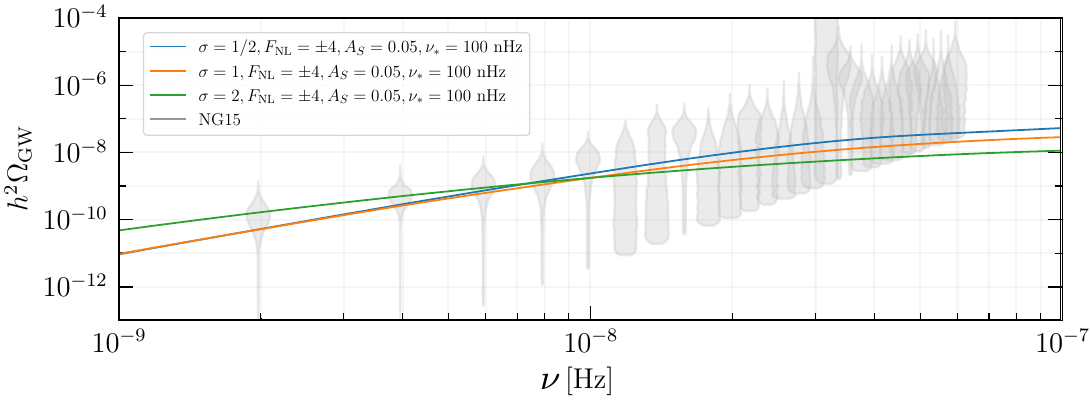} 
    \caption{Energy-density fraction spectra of \acp{SIGW} for different sets of independent parameters. The NG15 data are also shown for comparison. }
    \label{fig:omega_data}
\end{figure}

The energy-density fraction spectrum of \acp{SIGW} at the present conformal time $\eta_0$ can be expressed as
\begin{equation}\label{eq:ogwetaa0qsai} \bar{\Omega}_{\uGW,0} (\nu)  =  \Omega_{\mathrm{rad}, 0} \left[\frac{g_{\ast,\rho}(T)}{g_{\ast,\rho}(T_\mathrm{eq})} \right] \left[\frac{g_{\ast,s}(T_\mathrm{eq})}{g_{\ast,s}(T)} \right]^{4/3} \bar{\Omega}_\uGW (\eta,q) \ . \end{equation}
In the above equation, $\Omega_{\mathrm{rad}, 0}h^{2}=4.2 \times 10^{-5}$ represents the physical energy-density fraction of radiations in the present universe \cite{Planck:2018vyg}. 
$T$ and $T_\mathrm{eq}$ correspond to the cosmic temperatures at the emission time and the epoch of matter-radiation equality, respectively. 
$\nu$ can be related to $T$, $g_{\ast,\rho}(T)$, and $g_{\ast,s}(T)$ as follows \cite{Zhao:2022kvz}
\begin{eqnarray} \frac{\nu}{\mathrm{nHz}} = 26.5 \left(\frac{T}{\mathrm{GeV}}\right)\left(\frac{g_{\ast,\rho}(T)}{106.75}\right)^{1/2}\left(\frac{g_{\ast,s}(T)}{106.75}\right)^{-1/3}\ . \end{eqnarray}
Here, $g_{\ast,\rho}$ and $g_{\ast,s}$ represent the effective relativistic degrees of freedom in the universe, which are tabulated functions of $T$ as provided in Ref.~\cite{Saikawa:2018rcs}.
To illustrate the interpretation of current \ac{PTA} data in the framework of \acp{SIGW}, we depict $\bar{\Omega}_{\uGW,0}(\nu)$ with respect to $\nu$ in \cref{fig:omega_data}, using three specific sets of model parameters.

\section{Implications of PTA data for new physics}\label{sec:pps}

In this section, we investigate the potential constraints on the parameter space of the primordial power spectrum and \acp{PBH} using the \ac{NG15} data. While it is possible to obtain constraints from other \ac{PTA} datasets using the same methodology, we do not consider them in this study, as they would not significantly alter the leading results of our current work.

\begin{figure}
    \centering
    \includegraphics[width =1 \columnwidth]{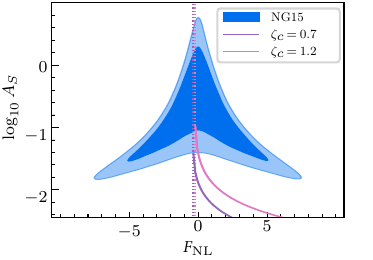} 
    \caption{Two-dimensional contours (blue shaded) in $\log_{10}A_{S}-\Fnl$ plane inferred from the NG15 data. Dotted lines denote $\Fnl=-(4\zeta_c)^{-1}$ while solid curves stand for models, which expect $f_{\mathrm{pbh}}=1$ for $m_{\mathrm{pbh}}=10^{-2}M_\odot$, in the cases of $\zeta_c=0.7$ (purple) and $\zeta_c=1.2$ (rose), respectively. }
    \label{fig:As_FNL}
\end{figure}

By performing a comprehensive Bayesian analysis \cite{NANOGrav:2023hvm}, we could gain valuable insights for the posteriors of four independent parameters, i.e., $\Fnl$, $A_S$, $\sigma$, and $\nu_\ast$, for which the priors are set to be $\Fnl\in[-30,30]$, $\log_{10}A_S \in [-3,1]$, $\sigma \in [0,5]$, and $\log_{10}(\nu_\ast/\mathrm{Hz}) \in [-9,-5]$. 
Here, we also adopt the aforementioned condition of perturbativity, namely, $A_{S}\Fnl^{2}<1$. 
The inference results within 68\% confidence intervals are given as 
\begin{eqnarray}
    \Fnl &=& -0.00^{+2.45}_{-2.46} \ ,\label{eq:fnl}\\
    \log_{10} A_{S} &=& -0.97^{+0.65}_{-0.46}\ ,\label{eq:as}\\
    \sigma &=& 1.08^{+1.08}_{-0.83}\ ,\label{eq:sigma}\\
    \log_{10}(\nu_{\ast}/\mathrm{Hz}) &=& -6.99^{+0.93}_{-0.45}\ .\label{eq:nuast}
\end{eqnarray} 
We can also recast Eq.~(\ref{eq:fnl}) into constraints on $\fnl$, i.e.,  
\begin{equation}
    \fnl = -0.0 \pm 4.1 \ . 
\end{equation}
Fig.~\ref{fig:As_FNL} shows two-dimensional contours in $\log_{10}A_{S}-\Fnl$ plane at 68\% (dark blue regions) and 95\% (light blue regions) confidence levels. 
There is a full degeneracy in the sign of primordial non-Gaussianity $\fnl$, as the energy-density fraction spectrum is dependent of only the absolute value of $\Fnl$, as demonstrated in Fig.~\ref{fig:unscaled_omega}.
The above results indicate that the \ac{PTA} observations have already emerged as a powerful tool for probing physics of the early universe.

We can further recast the constraints on the primordial curvature power spectrum into constraints on the nature of \acp{PBH}, which is characterized by their mass function. 
Due to significant uncertainties in the formation scenarios of \acp{PBH} (as discussed in reviews such as Ref.~\cite{Carr:2020gox}), we adopt a simplified scenario \cite{Meng:2022ixx} to illustrate the importance of primordial non-Gaussianity $\fnl$. 
The initial mass function of \acp{PBH} is described by 
\begin{equation}\label{eq:betang} \beta = \int_{\zeta>\zeta_c} P(\zeta) d\zeta = \int_{\zeta(\zeta_g)>\zeta_c}\frac{1}{\sqrt{2\pi} \sigma_g}\mathrm{exp}\left(-\frac{\zeta_g^2}{2\sigma_g^2}\right)d\zeta_g \ , \end{equation}
where $P(\zeta)$ represents the \ac{PDF} of primordial curvature perturbations, $\sigma_g$ is the standard variance of the Gaussian component $\zeta_g$ in the \ac{PDF}, and $\zeta_c$ stands for the critical fluctuation. 
We further find $\sigma_g^2=\langle \zeta_g^2\rangle=\int\Delta_g^2(q)d\ln q=A_S$ by considering the power spectrum defined in Eq.~(\ref{eq:Lognormal}). 
Additionally, it is known that $\zeta_c$ is of order $\mathcal{O}(1)$, with specific values of $0.7$ and $1.2$, as suggested by Ref.~\cite{Green:2004wb}. 

To evaluate Eq.~(\ref{eq:betang}), we devide $\Fnl$ into two regimes, i.e., $\Fnl>0$ and $\Fnl<0$. 
In the case of $\Fnl>0$, we solve the equation $\zeta(\zeta_g)=\zeta_c$, yielding a relation 
\begin{equation} \label{eq:saizgpm1}
\zeta_{g\pm} = \frac{-1\pm\sqrt{1+4\Fnl\zeta_c}}{2\Fnl}\ . \end{equation}
By substituting it into Eq.~(\ref{eq:betang}), we gain   
\begin{eqnarray} \beta &=& \left(\int_{-\infty}^{\zeta_{g-}} + \int^{+\infty}_{\zeta_{g+}}\right) P(\zeta_g)d\zeta_g \nonumber\\
&=& \frac{1}{2}\mathrm{erfc}\left(\frac{\zeta_{g+}}{\sqrt{2A_S}}\right) + \frac{1}{2}\mathrm{erfc}\left(-\frac{\zeta_{g-}}{\sqrt{2A_S}}\right)\ , \end{eqnarray}
where erfc(x) is the complementary error function. 
Similarly, in the case of $-(4\zeta_c)^{-1}<\Fnl<0$, we gain 
\begin{equation} \beta = \int_{\zeta_{g+}}^{\zeta_{g-}} P(\zeta_{g}) d\zeta_{g} = \frac{1}{2}\mathrm{erfc}\left(\frac{\zeta_{g+}}{\sqrt{2A_S}}\right) - \frac{1}{2}\mathrm{erfc}\left(\frac{\zeta_{g-}}{\sqrt{2A_S}}\right) \ . \end{equation}
In contrast, in the case of $\Fnl<-(4\zeta_c)^{-1}$, no \acp{PBH} were formed in the early universe, since the curvature perturbations are expected to never exceed the critical fluctuation. 
As a viable candidate for cold dark matter, the abundance of \acp{PBH} is determined as \cite{Nakama:2016gzw}
\begin{equation} f_{\mathrm{pbh}} \simeq 2.5\times10^{8}\beta\left(\frac{g_{\ast,\rho}(T_{\mathrm{f}})}{10.75}\right)^{-1/4}\left(\frac{m_{\mathrm{pbh}}}{M_\odot}\right)^{-1/2}\ , \end{equation}
where $m_{\mathrm{pbh}}$ represents the mass of \acp{PBH}, and $T_{\mathrm{f}}$ denotes cosmic temperature at the formation occasion. 
Roughly speaking, $m_{\mathrm{pbh}}$ can be related to the horizon mass $m_H$ and then the peak frequency $\nu_\ast$, namely, \cite{Kohri:2020qqd}
\begin{equation} \frac{m_{\mathrm{pbh}}}{M_\odot} \simeq \frac{m_H}{0.31M_\odot} \simeq \left(\frac{\nu_\ast}{5.0\mathrm{nHz}}\right)^{-2} \ . \end{equation}

Based on Eq.~(\ref{eq:nuast}), we could infer that the mass range of \acp{PBH} is the order of $\mathcal{O}(10^{-5}-10^{-2})M_\odot$. 
However, the inferred abundance of \acp{PBH} exceeds unity in the case of a sizable positive $\Fnl$, indicating an overproduction of \acp{PBH}. 
This is because the inferred value of $A_S$ is typically one order of magnitude larger than the value of $A_S$ that leads to $f_{\mathrm{pbh}}=1$.
To illustrate this result more clearly, we include into Fig.~\ref{fig:As_FNL} two solid curves corresponding to $m_{\mathrm{pbh}}=10^{-2}M_\odot$ and $f_{\mathrm{pbh}}=1$ in the cases of $\zeta_c=0.7$ (purple curve) and $\zeta_c=1.2$ (rose curve), respectively. 
For comparison, we mark the critical value $\Fnl=-(4\zeta_c)^{-1}$ with dotted lines. 
Therefore, we find that a negative $\Fnl$ is capable of alleviating the overproduction of \acp{PBH}, especially when considering a sizable negative $\Fnl$, namely, $\Fnl<-(4\zeta_c)^{-1}$, which prevents the formation of any \acp{PBH}. 
However, due to large uncertainties in model buildings, it remains challenging to exclude the \ac{PBH} scenario through analyzing the present \ac{PTA} data. 

In summary, it is crucial to measure the primordial non-Gaussianity or at least determine the sign of $\Fnl$ in order to assess the viability of the \ac{PBH} scenario. 
However, it is impossible to determine the sign of $\Fnl$ through measurements of the energy-density fraction spectrum of \acp{SIGW}, due to the sign degeneracy. 
In the next section, we will propose that the inhomogeneous and anisotropic component of \acp{SIGW} has the potential to break the sign degeneracy, as well as other degeneracies in model parameters, opening up new possibilities for making judgments about the \ac{PBH} scenario in the future.

\section{SIGW angular power spectrum}\label{sec:asigw}

In this section, we investigate the inhomogeneities and anisotropies in \acp{SIGW} via deriving the angular power spectrum in the \ac{PTA} band, following the research approach established in our previous paper \cite{Li:2023qua}.

The inhomogeneities in \acp{SIGW} arise from the long-wavelength modulations of the energy density generated by short-wavelength modes. 
As discussed in Section \ref{sec:edfs}, \acp{SIGW} originate from extremely high redshifts, corresponding to very small horizons. 
However, due to limitations in the angular resolution of detectors, the signal along a line-of-sight represents an ensemble average of the energy densities over a sizable number of such horizons. 
Consequently, any two signals would appear identical. 
Nevertheless, the energy density of \acp{SIGW} produced by short-wavelength modes can be spatially redistributed by long-wavelength modes if there are couplings between the two. 
The local-type primordial non-Gaussianity $\fnl$ could contribute to such couplings.

Similar to the temperature fluctuations of relic photons \cite{Seljak:1996is}, the initial inhomogeneities in \acp{SIGW} at a spatial location $\bx$ can be characterized by the density contrast, which is denoted as $\delta_\uGW (\eta,\bx,\bq)$, given by \begin{equation} \delta_\uGW (\eta,\bx,\bq) = 4\pi \frac{\omega_\uGW (\eta,\bx,\bq)}{\bar{\Omega}_\uGW (\eta,q)} - 1, \end{equation} 
where the energy-density full spectrum $\omega_{\uGW}(\eta,\bx,\bq)$ is defined in terms of the energy density, namely, $\rho_\uGW (\eta,\bx) = \rho_\uc(\eta) \int \ud^3 \bq, \omega_\uGW (\eta,\bx,\bq)/q^3$. 
We specifically get $\omega_\uGW \sim \langle\zeta^4\rangle_{\bx}$, where the subscript $_\bx$ denotes an ensemble average within the horizon enclosing $\bx$ \cite{Bartolo:2019zvb,Li:2023qua}.
We decompose $\zeta_g$ into modes of short-wavelength $\zeta_{gS}$ and long-wavelength $\zeta_{gL}$, namely, $\zeta_g=\zeta_{gS}+\zeta_{gL}$ \cite{Tada:2015noa}. 
At linear order in $\zeta_{gL}$, we get $\delta_\uGW \sim \zeta_{gL} \langle\zeta_{gS}\zeta_{S}^{3}\rangle_{\bx}$, where $\zeta_{S}$ represents the part of $\zeta$ composed solely of $\zeta_{gS}$. 
Terms of higher orders in $\zeta_{gL}$ are negligible due to smallness of the power spectrum $\Delta_L^2\sim10^{-9}$ \cite{Planck:2018vyg}. 
Using Feynman-like rules and diagrams, we get an expression for $\delta_\uGW(\eta,\bx,\bq)$, i.e., \cite{Li:2023qua} 
\begin{equation} \delta_{\uGW}(\eta,\bx,\bq) = \Fnl \frac{\Omega_{\mathrm{ng}}(\eta,q)}{\bar{\Omega}_{\uGW}(\eta,q)} \int \frac{d^{3}\bk}{(2\pi)^{3/2}} e^{i\bk\cdot\bx} \zeta_{gL}(\bk)\ , \end{equation} 
where we introduce a quantity of the form 
\begin{equation} 
\Omega_{\mathrm{ng}} (\eta,q) = 2^3 \bar{\Omega}_{\uGW}^{(0)} (\eta,q) + 2^2 \bar{\Omega}_\uGW^{(1)} (\eta,q)\ . 
\end{equation} 

The present density contrast, denoted as $\delta_{\uGW,0}(\bq)$, can be estimated analytically using the line-of-sight approach \cite{Contaldi:2016koz,Bartolo:2019oiq,Bartolo:2019yeu}. 
It is contributed by both the initial inhomogeneities and propagation effects, given by \cite{Bartolo:2019zvb} 
\begin{equation} 
\delta_{\uGW,0}(\bq) = \delta_\uGW (\eta,\bx,\bq) + \left[4-n_{\uGW,0} (\nu)\right] \Phi (\eta, \bx)\ . 
\end{equation} 
Here, $n_{\uGW,0} (\bq)$ denotes the index of the present energy-density fraction spectrum in Eq.~(\ref{eq:ogwetaa0qsai}), given by 
\begin{equation} 
n_{\uGW,0} (\nu) = \frac{\partial\ln \bar{\Omega}_{\uGW,0} (\nu)}{\partial\ln \nu} \simeq \frac{\partial\ln \bar{\Omega}_{\uGW} (\eta,q)}{\partial\ln q}\Big|_{q=2\pi\nu}\ . 
\end{equation} 
For the propagation effects, we consider only the \ac{SW} effect \cite{Sachs:1967er}, which is characterized by the Bardeen's potential on large scales 
\begin{equation} 
\Phi (\eta,\bx) = \frac{3}{5} \int \frac{d^{3}\bk}{(2\pi)^{3/2}} e^{i\bk\cdot\bx} \zeta_{gL}(\bk)\ . 
\end{equation} 
We assume the statistical homogeneity and isotropy for the density contrasts on large scales, similar to the study of \ac{CMB} \cite{Maggiore:2018sht}. 

The anisotropies today can be mapped from the aforementioned inhomogeneities. 
The reduced angular power spectrum is useful to characterize the statistics of these anisotropies. 
It is defined as the two-point correlator of the present density contrast, namely, 
\begin{equation} 
\langle\delta_{\uGW,0,\ell m}(2\pi\nu) \delta_{\uGW,0,\ell' m'}^\ast(2\pi\nu)\rangle = \delta_{\ell \ell'} \delta_{mm'} \widetilde{C}_{\ell} (\nu)\ , 
\end{equation} 
where $\delta_{\uGW,0}(\bq)$ has been expanded in terms of spherical harmonics, i.e., 
\begin{equation}
    \delta_{\uGW,0}(\bq) = \sum_{\ell m} \delta_{\uGW,0,\ell m}(q) Y_{\ell m}(\bn) \ .
\end{equation} 
Roughly speaking, we get $\widetilde{C}_{\ell} \sim \delta_{\uGW,0}^{2} \propto \langle \zeta_{gL}\zeta_{gL} \rangle \sim \Delta_{L}^{2}$. 
Detailed analysis using Feynman-like rules and diagrams was conducted in our previous paper \cite{Li:2023qua}. 
We summarize the main results as follows \begin{equation}\label{eq:reduced-APS} \widetilde{C}_{\ell} (\nu) = \frac{18\pi\Delta^2_L}{25 \ell (\ell+1)} \biggl\{ \fnl \frac{\Omega_{\mathrm{ng}} (\eta,2\pi\nu)}{\bar{\Omega}_\uGW (\eta,2\pi\nu)} + \bigl[4 - n_{\uGW,0} (\nu)\bigr] \biggr\}^2\ , \end{equation} which can be recast into the angular power spectrum 
\begin{equation} 
C_{\ell}(\nu) = \left[\frac{\bar{\Omega}_{\uGW,0}(\nu)}{4\pi}\right]^{2} \tilde{C}{\ell}(\nu)\ . 
\end{equation} 
Analogous to \ac{CMB}, for which the root-mean-square (rms) temperature fluctuations is determined by $[{\ell(\ell+1)C^{\mathrm{CMB}}_{\ell}/(2\pi)}]^{1/2}$, the rms density contrast for \acp{SIGW} is determined by $[{\ell(\ell+1)C_{\ell}(\nu)/(2\pi)}]^{1/2}$, which represents the variance of the energy-density fluctuations. 
It is vital to note that the rms density contrast is constant with respect to multipoles $\ell$, but depends on frequency bands.

\begin{figure}
    \centering
    \includegraphics[width =1 \columnwidth]{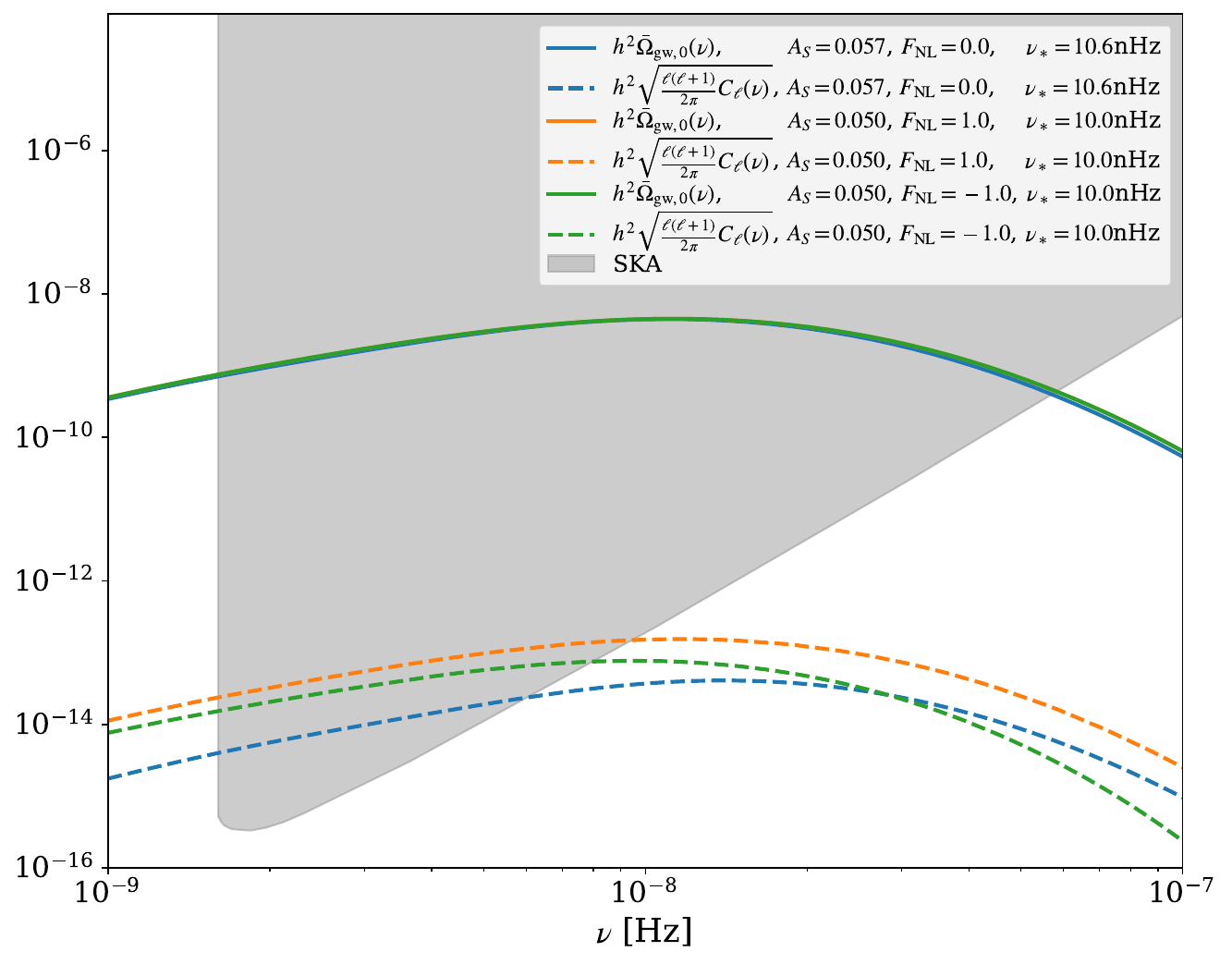}
    \caption{Physical energy-density fraction spectra of SIGWs, $h^2 \Omega_{\uGW,0} (\nu)$ (solid), and the variance of SIGW density contrasts, $h^2[{\ell(\ell+1)C_{\ell}(\nu)/(2\pi)}]^{1/2}$ (dashed). For comparison, we show the sensitivity region of SKA \cite{Schmitz:2020syl} (gray shaded). }
    \label{fig:delta_omega}
\end{figure}

In Figure \ref{fig:delta_omega}, we present the rms density contrast as a function of gravitational-wave frequency. 
We also include the energy-density fraction spectrum for comparison. 
Roughly speaking, we find that $\sqrt{\tilde{C}_{\ell}}$ is the order of $10^{-4}$, depending on specific sets of model parameters. 
It is worth noting that the angular power spectrum can break degeneracies among these parameters. 
For instance, based on Fig.~\ref{fig:delta_omega}, we observe a coincidence in the energy-density fraction spectra for three different parameter sets. 
However, the angular power spectrum breaks this coincidence, particularly in the case of the sign degeneracy of $\fnl$. 
This result suggests that measurements of the anisotropies in \acp{SIGW} have the potential to determine the primordial non-Gaussianity \cite{Li:2023qua}.
Recently, an upper limit of $\tilde{C}_{\ell}<20\%$ was inferred from the \ac{NG15} data \cite{NANOGrav:2023tcn}. 
However, this limit is not precise enough to test the theoretical predictions of our present work. 
In contrast, based on Fig.~\ref{fig:delta_omega}, we anticipate that the \ac{SKA} program \cite{Schmitz:2020syl} will offer sufficient precision to measure the non-Gaussianity $\fnl$.

\section{Conclusions}\label{sec:cd}

In this study, we examined the implications of recent \ac{PTA} datasets for understanding the nature of primordial curvature perturbations and primordial black holes (\acp{PBH}). 
Specifically, we investigated the influence of primordial non-Gaussianity $\fnl$ on the inference of model parameters, and vice versa, by analyzing the recent \ac{NG15} data.
In particular, at 68\% confidence level, we inferred $|\fnl|<4.1$, which is competitive with the constraints from measurements of \ac{CMB}. 
Even when considering the non-Gaussianity $\fnl$, we found that the \ac{PBH} scenario is in tension with the \ac{NG15} data, except when a sizable negative $\fnl$ is considered, which can significantly suppress the abundance of \acp{PBH}. 
Our results indicated that the \ac{PTA} observations have already emerged as a powerful tool for probing physics of the early universe and dark matter. 
Moreover, we proposed that the anisotropies of \acp{SIGW} serve as a powerful probe of the non-Gaussianity $\fnl$ in the \ac{PTA} band. 
For the first time, we conducted the complete analysis of the angular power spectrum in this frequency band and found that it can effectively break potential degeneracies among the model parameters, particularly the sign degeneracy of $\fnl$. 
Additionally, we explored the detectability of the anisotropies in \acp{SIGW} in the era of the \ac{SKA} project.

\vspace{0.3cm}
\hspace{0.2cm}
\emph{Notes added.}--- 
During the preparation of this paper, a related study \cite{Franciolini:2023pbf} appears, which examines the posteriors of \ac{NG15} data. The authors suggest that the Gaussian scenarios for \acp{SIGW} are in tension with the current \ac{PTA} data at a $2\sigma$ confidence level, but non-Gaussian scenarios that suppress the abundance of \acp{PBH} can alleviate this tension. Given the significant uncertainties in the formation scenarios of \acp{PBH} (as discussed in reviews such as Ref.~\cite{Carr:2020gox}), the main focus of our research is to simultaneously examine the energy-density fraction spectrum and the angular power spectrum of \acp{SIGW}, by incorporating the complete contributions arising from primordial non-Gaussianity $\fnl$. We also address the importance of primordial non-Gaussianity to \acp{SIGW} through a Bayesian analysis over the \ac{NG15} data.

\begin{acknowledgments}
S.W. and J.P.L. are supported by the National Natural Science Foundation of China (Grant NO. 12175243) and the Science Research Grants from the China Manned Space Project with No. CMS-CSST-2021-B01. Z.C.Z. is supported by the National Natural Science Foundation of China (Grant NO. 12005016).  Z.Q.Z is supported by the National Natural Science Foundation of China (Grant NO. 12305073).
\end{acknowledgments}

\appendix

\begin{widetext}
\section{Formulae for evaluating the SIGW energy density}\label{sec:formula}

After a comprehensive derivation following the methodology presented in Refs.~\cite{Adshead:2021hnm,Li:2023qua,Ragavendra:2021qdu}, we can precisely express the three terms in Eq.~(\ref{eq:Omegabar-total}) as  
\begin{eqnarray}
    \bar{\Omega}_\uGW^{(0)} (\eta, q) 
    &=& \frac{1}{3} \int_0^\infty \ud t_1 \int_{-1}^1 \ud s_1 
        \overbar{J^2 (u_1,v_1,x\rightarrow\infty)} \frac{1}{(u_1 v_1)^2} 
        \Delta^2_g (v_1 q) \Delta^2_g (u_1 q) \ ,\label{eq:Omega-G}\\
    \bar{\Omega}_\uGW^{(1)} (\eta,q) 
    &=& \frac{\Fnl^2}{3\pi} 
        \prod_{i=1}^2 \biggl[\int_0^\infty \ud t_i \int_{-1}^1 \ud s_i\, v_i u_i\biggr] 
        \Bigg\{ \frac{\pi \overbar{J^2 (u_1,v_1,x\rightarrow\infty)}}{(u_1 v_1 u_2 v_2)^3} \Delta^2_g (v_1 v_2 q) \Delta^2_g (u_1 q) \Delta^2_g (v_1 u_2 q)\nonumber\\
        &&\qquad+ \int_0^{2\pi} \ud \varphi_{12}\, \cos 2\varphi_{12} \overbar{J(u_1,v_1,x\rightarrow\infty)J(u_2,v_2,x\rightarrow\infty)} \label{eq:Omega-C}\\
        &&\hphantom{\qquad\ + \int_0^{2\pi} \ud \varphi_{12}}
        \times \frac{\Delta^2_g (v_2 q)}{v_2^3} \frac{\Delta^2_g (w_{12} q)}{w_{12}^3} \bigg[\frac{\Delta^2_g (u_2 q)}{u_2^3} + \frac{\Delta^2_g (u_1 q)}{u_1^3} \bigg] \Bigg\}
            \ ,\nonumber\\
    \bar{\Omega}_\uGW^{(2)} (\eta,q) 
    &=& \frac{\Fnl^4}{24\pi^2} 
        \prod_{i=1}^3 \biggl[\int_0^\infty \ud t_i \int_{-1}^1 \ud s_i\, v_i u_i\biggr] \Bigg\{ \frac{2\pi^2 \overbar{J^2 (u_1,v_1,x\rightarrow\infty)}}{(u_1 v_1 u_2 v_2 u_3 v_3)^3} \Delta^2_g (v_1 v_2 q) \Delta^2_g (v_1 u_2 q) \Delta^2_g (u_1 v_3 q) \Delta^2_g (u_1 u_3 q)\nonumber\\
    &&\qquad\ + \int_0^{2\pi} \ud \varphi_{12}\ud \varphi_{23}\, \cos 2\varphi_{12} \overbar{J (u_1,v_1,x\rightarrow\infty) J (u_2,v_2,x\rightarrow\infty)} \label{eq:Omega-N}\\ 
        &&\hphantom{\qquad\ + \int_0^{2\pi} \ud \varphi_{12}\ud \varphi_{23}\,} 
        \times \frac{\Delta^2_g (u_3 q)}{u_3^3} \frac{\Delta^2_g (w_{13} q)}{w_{13}^3} 
            \bigg[\frac{\Delta^2_g (v_3 q)}{v_3^3}  \frac{\Delta^2_g (w_{23} q)}{w_{23}^3} + \frac{\Delta^2_g (w_{23} q)}{w_{23}^3}
            \frac{\Delta^2_g (w_{123} q)}{w_{123}^3}\bigg]\Bigg\} \ , \nonumber 
\end{eqnarray}
where we define $x=q\eta$, $s_i=u_i-v_i$, $t_i=u_i+v_i-1$, and 
\begin{subequations}
\begin{eqnarray}
    y_{ij}&=&\frac{\cos\varphi_{ij}}{4}\sqrt{t_i(t_i+2)(1-s_i^2)t_j(t_j+2)(1-s_j^2)}+\frac{1}{4}[1-s_i(t_i+1)][1-s_j(t_j+1)]\ , \\
    w_{ij}&=&\sqrt{v_i^2+v_j^2-y_{ij}}\ ,\\
    w_{123}&=&\sqrt{v_1^2+v_2^2+v_3^2+y_{12}-y_{13}-y_{23}}\ .
\end{eqnarray}
\end{subequations}
The calculation for the average of the squared oscillation $J(u,v,x\rightarrow\infty)$ has been provided in Ref.~\cite{Li:2023qua}, as well as in earlier studies referenced in Refs.~\cite{Espinosa:2018eve,Kohri:2018awv,Atal:2021jyo,Adshead:2021hnm}, i.e., 
\begin{eqnarray}\label{eq:J-ave-12}
    &&\overbar{J (u_i,v_i,x\rightarrow\infty)J (u_j,v_j,x\rightarrow\infty)}\nonumber\\ 
    &=&\frac{9 \left(1-s_i^2\right) \left(1-s_j^2\right) t_i \left(t_i+2\right) t_j \left(t_j+2\right) \left(s_i^2+t_i^2+2 t_i-5\right) \left(s_j^2+t_j^2+2 t_j-5\right) }{8 \left(-s_i+t_i+1\right){}^3 \left(s_i+t_i+1\right){}^3 \left(-s_j+t_j+1\right){}^3 \left(s_j+t_j+1\right){}^3}\nonumber\\
    &&\times \Bigg\{\left[\left(s_i^2+t_i^2+2 t_i-5\right) \ln \left(\left| \frac{t_i^2+2 t_i-2}{s_i^2-3}\right| \right) + 2\left(s_i-t_i-1\right) \left(s_i+t_i+1\right)\right] \nonumber\\
    &&\hphantom{\times \Bigg\{} \times \left[\left(s_j^2+t_j^2+2 t_j-5\right) \ln \left(\left| \frac{t_j^2+2 t_j-2}{s_j^2-3}\right| \right) + 2\left(s_j-t_j-1\right) \left(s_j+t_j+1\right)\right]\\
    &&\hphantom{\times \Bigg\{}+\pi^2\Theta\left(t_i-\sqrt{3}+1\right) \Theta\left(t_j-\sqrt{3}+1\right) \left(s_i^2+t_i^2+2 t_i-5\right) \left(s_j^2+t_j^2+2 t_j-5\right)\Bigg\}\ .\nonumber 
\end{eqnarray}
The equations presented in this appendix can be utilized to numerically calculate the energy density of \acp{SIGW} in a self-consistent manner. 

\end{widetext}

\bibliography{biblio.bib}

\end{document}